\documentclass[%
preprint,
 amsmath,amssymb,aps,prl
]{revtex4-1}

\usepackage{graphicx}
\usepackage{dcolumn}
\usepackage{bm}

\begin{document}

\title{Cell Migration Model with Multiple Chemical Compasses}
\author{Shuji Ishihara}
 \email[E-mail address: ]{shuji@complex.c.u-tokyo.ac.jp}
\affiliation{Graduate School of Arts and Sciences, University of Tokyo, Meguro-ku, Tokyo 153-8902, Japan}

\begin{abstract}
    A simple model is proposed that describes the various
    morphodynamic principles of migrating cells from polar to
    amoeboidal motions. The model equation is derived using competing
    internal cellular compass variables and symmetries of the
    system. Fixed points for the $N = 2$ system are closely
    investigated to clarify how the competition among polaritors
    explains the observed morphodynamics. Response behaviors of cell--to--signal stimuli are also investigated. This model will be useful
    for classifying high-dimensional cell motions and investigating
    collective cellular behaviors.
\end{abstract}

\pacs{87.17.Jj,87.17.Aa,05.45.Jn}

\maketitle


The migration of cells on substrates is a key component of various
biological functions, such as development, immune system response, and
wound healing \cite{Bray}. Cells exhibit diverse and involved
morphodynamics depending on their type, developmental stage, and
environmental conditions, and yet represent ordered and common
dynamics across a range of species
\cite{RidleySci2003,MoglinerCurrBiol2009}. Many cells exhibit an
apparent monopolar shape, consisting of a head and a tail.  The
keratocyte-like crescent shape is also widely observed in nature
\cite{KerenNat2008}.  The highly investigated \textit{Dictyosterium}
cells form a roundish shape with amoeboidal, non-directed, random
protrusions. Under poor nutrient conditions, these cells become
elongated with a definitive polarity and exhibit a zig-zag
(``split-and-choice'') motion \cite{AndrewNCB2007,Otsuji2010},
followed by a collective spiral migration pattern \cite{Sawai}.  These
cellular behaviors are regulated by the cytoskeleton (specifically,
actin filaments) mediated by signaling molecules (\textit{e.g.},
phosphoinositide lipids).  It is now recognized that interactions
among these molecules lead to the instability of a uniform molecular
distribution inside a cell and generate self-organized chemical waves
to support complex cell morphodynamic processes
\cite{Vicker,Gerisch,Schoth-DiezHSFP2009,WeinerPLoSBiol2007,Arai20102012,Taniguchi}.
A number of theoretical models have been proposed by considering
associated chemical reactions
\cite{Otsuji2010,WeinerPLoSBiol2007,Arai20102012,Taniguchi,Jinken,Otsuji2007,Skpsky,Shao,Neilson2011}
or actin polymerization
\cite{Nishimura,ShlomovitzPRL2007,Carlsson2010,DoubrovinskiPRL2011,EnculescuNJP2011,ZIebertJRSI2012}.
However, these models are primarily focused on the onset of
instabilities and a relatively simple pattern, and not generally
intended to elucidate how a variety of morphodynamic processes from
ordered to amoeboidal cell motions are organized. One of difficulties
in addressing this issue is the requirement of large computational
power for the execution of these models.  Thus, simplified modeling
with appropriate abstraction is another theoretical challenge in
identifying the mechanism of diversity in cellular morphodynamics.

\begin{figure}[bt]
\centering
  \resizebox{0.45\textwidth}{!}{%
    \includegraphics{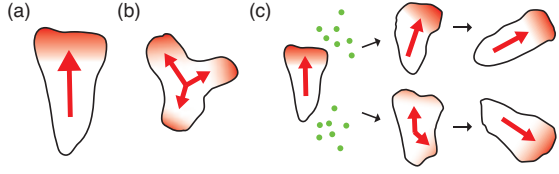}
  }
\caption{\label{fig:Fig1} (a) Polarized cell represented by a single
  cellular compass variable (polaritor) (b) Amoeboidal cell
  represented by multiple polaritors. (c) Distinct responses to signal
  stimuli from different directions. The existing cell head turns in
  response to stimuli from the front to change direction (top),
  whereas the cell head is replaced by a new head in response to
  signals from the rear (bottom).}
\end{figure}

Here, we adopt one of the familiar concepts known as a ``chemical
compass," which was introduced as a hypothetical internal cellular
state and an intuitive representation of intrinsic cell directionality
(Fig.~\ref{fig:Fig1}(a))
\cite{RickertTCB2000,BourneNature2002,MeiliCell2003,SunBool2004,ArrieumerlouDevCell2005,KingTreCellBiol2009}. The
compass can be interpreted to represent cellular polarity as dictated
by molecular distribution. Based on the potential high dimensionality
of molecular dynamics, there may be multiple cell compasses rather
than only a single one. In practice, this is evident in amoeboidal
motion, which exhibits a number of protrusions with patched molecular
localization at the cell boundary (Fig.~\ref{fig:Fig1}(b)). This is
also supported by the observed mechanisms by which a cell responds to
signal stimuli from different directions; cells turn by rotating their
existing head in response to signals, although the existing head
occasionally disappears in response to signals from the rear and a new
head is formed in the direction of the signal (Fig.~\ref{fig:Fig1}(c))
\cite{AndrewNCB2007}. These observations motivated us to introduce
multiple compass variables for describing cellular dynamics. A model
is derived in the following text for single cellular motion based on
the competitive dynamics occurring among the hypothetical compass
variables.

Intrinsic compass variables in the cell are represented by complex
variables $W_i = R_i e^{{\rm i} \theta_i} (i=1,2,\ldots,N)$ with their
amplitudes ($R_i$) and directions ($\theta_i$).  We referr to the
variables as polaritors in this study.  A part of polaritors is
supposed to be simultaneously active (\textit{i.e.}, $R_i > 0$), while
the other polaritors are in a quiescent state (\textit{i.e.}, $R_i =
0$).  The dynamics of these polaritors should satisfy the following
requirements: First, a single polaritor spontaneously breaks
$\mbox{U(1)}$ symmetry, thereby obeying an equation similar to
$\dot{W} = W - |W|^mW$ for an even integer $m$.  Second, the system
should satisfy all plausible symmetries; in other words, the equation
should be invariant by the following transformation of variables: (i)
simultaneous shift of polaritors' direction $\theta_i \to \theta_i +
\theta_0$ (isotropy), (ii) reflection $\theta_i \to - \theta_i$ for
all $i$ (mirror symmetry), and (iii) permutation among polaritors'
indices. Third, the system should possess invariant subspaces
specified by $W_i = 0$ for each $i$, since polaritors in quiescent
states should not influence the dynamics of the system. On the basis
of these requirements, the following equation is proposed for
the dynamics of $W_i$ \cite{Caption1};
\begin{eqnarray}
    \dot{W}_i \!&=&\! W_i \!-\! |W_i|^4W_i \!-\!\gamma \left( \Sigma' |W_j|^2\right)
    W_i \!-\! \delta \left( \Sigma' W_j^2\right)\overline{W}_i ~~~ \nonumber \\
    \!&&\! +~ c_1 \left( \Sigma' W_j\right)|W_i|^2
    + c_2 \left( \Sigma' \overline{W}_j\right) W_i^2
    + \xi_i(t). \label{eq:Wequation}
\end{eqnarray}
In this equation, $\overline{W_i}$ is the complex conjugate of $W_i$, $\Sigma'
\equiv \Sigma_{j \neq i}$ represents the calculation of the sum over $j$ except $i$,
and the coefficients $\gamma, \delta, c_1$ and $c_2$ are real numbers
due to the requirement of symmetry (ii). $m=4$ is adopted to ensure
the stability of the system for arbitrary parameter values. The final
term $\xi_i(t)$ is added to incorporate white Gaussian noise into the statistics $\langle \xi_i(t)\rangle =0$ and $\langle \xi_i(t)
\xi_j(t')\rangle = 2D\delta_{ij}\delta(t-t')/N$.  In polar coordinates,
Eq.~(\ref{eq:Wequation}), without the noise term, is expressed as
\begin{eqnarray}
    \dot{R}_i \!&=&\! R_i-R_i^5
    - \Sigma' \left( \gamma+\delta \cos 2\!\left(\theta_j\!-\!\theta_i\right) \right)R_j^2R_i \nonumber \\
    & &\! +~ \mu \Sigma' \cos \left(\theta_j\!-\!\theta_i\right)
    R_jR_i^2, \label{eq:Rj} \\
    \dot{\theta}_i \!&=&\!
    -\delta \Sigma'R_j^2 \sin 2\!\left(\theta_j\!-\!\theta_i\right)
    + \nu \Sigma' \sin \left( \theta_j\!-\!\theta_i\right)\! R_jR_i  \label{eq:thetaj}
\end{eqnarray}
where $\mu = c_1+c_2$ and $\nu = c_1-c_2$. These equations indicate that
$\gamma$ represents angle-independent competition among polaritors, while $\delta$ and $\mu$ represent angle-dependent interactions of the first and second order. The velocity of the cell centroid  $\bf{x}$
is set to obey the equation
\begin{eqnarray}
    \dot{\textbf{x}} = \textbf{v} = \sum {}_{i} W_i.     \label{eq:Cvel}
\end{eqnarray}
For visibility, the shape of the cell is assumed to depend on polaritors as an angular-radius function  $L(\theta) = c^{-1}L_0(\theta)$ with $L_0(\theta) = \left(R_0+\sum_i R_i e^{\Lambda \cos(\theta - \theta_i)}\right) $ and with the normalization
factor $c^2 = \frac{1}{2A}\int_0^{2\pi}\!\! L_0^2(\theta) d \theta $ to
keep the cell area constant.

\begin{figure}[tb]
\centering
  \resizebox{0.49\textwidth}{!}{%
    \includegraphics{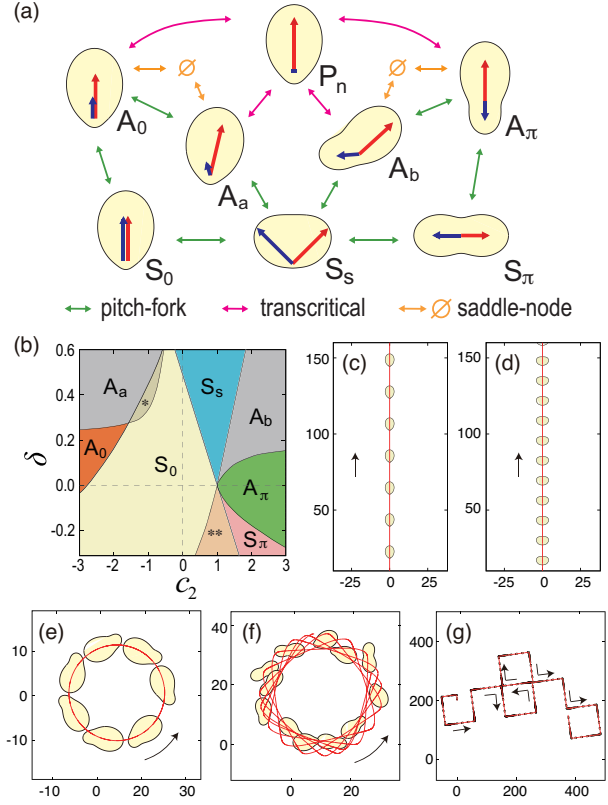}
  }
\caption{\label{fig:Fig2} (a) Fixed points of the $N=2$ system and
  their relationship (see text). Red and blue arrows indicate
  polaritors.  (b) Phase diagram of the $N=2$ system. $\gamma$ and
  $c_1$ are fixed as $\gamma = 0.7$ and $c_1=1.0$. Different types of
  fixed points appear, depending on the initial condition in regions
  indicated by $\ast$ ($\bf{A_0}$ and $\bf{S_0}$) and $\ast\ast$
  ($\bf{S_0}$ and $\bf{S_{\pi}}$). $\bf{P_n}$ may appear when $\gamma$
  is sufficiently large ($\gamma > 1+|\delta| $).  (c-f) Examples of
  numerical simulations for the $N=2$ system. Red lines indicate the
  trajectories of cell centroids.  Parameters $(\gamma, \delta, c_1,
  c_2 )$ are set as (c) $(1.2, 0.1, 1.0, 0.5)$, (d) $(0.7, 0.4, 1.0,
  1.0)$, (e) $(0.7,0.2,1.0,1.6)$, and (f) $(0.78,0.38,-0.45,0.45)$.
  Parameters for shape are set as $R_0 = 1.8$,$\Lambda = 2.4$, and
  $A=0.3$.}
\end{figure}

To understand how the competition among polaritors gives rise to new
dynamics, we first examined the simplest $N = 2$ system in the absence
of noise. With definitions of relative angle $\psi \equiv \theta_2
-\theta_1$ and mean angle $\Psi \equiv (\theta_1+\theta_2)/2$,
Eqs. (\ref{eq:Rj}) and (\ref{eq:thetaj}) reduce to
\begin{eqnarray}
    \dot{R}_1 \!&=&\! R_1\!-\!R_1^5 \!-\!  \left( \gamma+\delta \cos 2\psi \right)\!R_2^2R_1  +\mu  \cos  \psi R_2R_1^2,~~~ \label{eq:R1/2}\\
    \dot{R}_2 \!&=&\! R_2\!-\!R_2^5 \!-\!  \left( \gamma+\delta \cos 2\psi \right)\!R_1^2R_2 +\mu  \cos  \psi R_1R_2^2,~~~ \label{eq:R2/2}\\
    \dot{\psi} \!&=&\! \delta \left(R_1^2+R_2^2 \right)\sin 2 \psi -2\nu \sin \psi R_1R_2, \label{eq:Psi2} \\
    \dot{\Psi} \!&=&\! \delta (R_2^2-R_1^2) \sin 2\psi / 2.  \label{eq:EqCT}
\end{eqnarray}
The equations for $R_1, R_2$, and $\psi$ are incorporated within these
variables; $\Psi$ evolves depending on these variables, and thus, the
dynamics related to $(R_1, R_2, \psi)$ is the primary interest
herein. The domain of $\psi$ can be restricted to $-\pi \!<\!  \psi
\!\leq\! \pi $.  We observed that a variety of fixed points appear in
this system (Fig.~\ref{fig:Fig2}(a)) as listed below, where several
obvious fixed points obtained by replacing $i=1$ and $2$ are excluded.
The stability of these fixed points are also determined by considering
the linear equation $\dot{\vec{\rho}} = J \vec{\rho}$, where
$\vec{\rho} $ indicates a small deviation in $(R_1, R_2, \psi)$ and
$J$ is a Jacobian matrix at a fixed point. Eigenvalues and
corresponding eigenvectors of $J$ are denoted as $\lambda \!=\!
(\lambda_1, \lambda_2, \lambda_2)$ and
$\vec{\rho}_1,\vec{\rho}_2,\vec{\rho}_3$, respectively.
Fig.~\ref{fig:Fig2}(a) summarizes the fixed points and their
relationships.

\textbf{[$\bf{Z}$]} $(R_1, R_2) = (0,0)$ for arbitrary $\psi$. This
is a trivial fixed point with $\lambda_1 = \lambda_2 = 1$ and
$\lambda_3 =0$, and always remains unstable.

\textbf{[$\bf{P_n}$]} $(R_1,R_2, \psi) = (1,0, n \pi/2)$ with $n$ as
an integer (Fig.~\ref{fig:Fig2}(a)).  These fixed points are also an
expected trivial state, where one polaritor is active while the other
is quiescent. Eigenvalues are denoted as $\lambda_1 = -4, \lambda_2 =
1-(\gamma + \delta \cos 2\psi)$, and $\lambda_3 = 2 \delta \cos
2\psi$. One of these fixed points is stable as long as $\gamma >
1+|\delta|$.

\textbf[$\bf{S_0}, \bf{S_{\pi}}, \bf{S_{s}}$] Fixed points with
symmetric finite amplitudes ($R_1\!=\!R_2\!=R\!>\!0$;
Fig.~\ref{fig:Fig2}(a)).  Directions of the two polaritors coincide
for $\rm{\bf S_0}$ ($\psi\!=\!0$), but are reversed for $\rm{\bf
  S_{\pi}}$ ($\psi \!=\! \pi$). $\rm{\bf S_{s}}$ has a non-trivial
relative angle $\psi_s$ that is determined by $\cos \psi_s =
\nu/2\delta$.  Square amplitudes $R_{0}$, $R_{\pi}$, and $R_s$ are
defined by $R^2 = [ - (\gamma +\delta \cos 2 \psi -\mu \cos \psi ) +
\sqrt{(\gamma +\delta \cos 2 \psi -\mu \cos \psi )^2+4}]/2$ with the
substitution of $\psi = 0$, $\pi$, and $\psi_s$, respectively.  The
linear stability of these fixed points are evaluated by $\lambda_1 =
2-6R^4$, $\lambda_2 = -2 -2R^4$, and $\lambda_3 = 2R(2 \delta \cos
2\psi - \nu \cos \psi )$. The eigenvector of $\lambda_3$ is
$\vec{\rho}_3 = (0,0,1)$, indicating that $\lambda_3$ defines
stability in the direction of the relative angle $\psi$ (angular
stability).  $\rm{\bf S_{0}}$ and $\rm{\bf S_{\pi}}$ are unstable in
the angular direction in the regions $\nu < 2\delta$ and $\nu >
-2\delta$, respectively, in which $\rm{\bf S_{s}}$ appears through
pitch-fork bifurcation. In contrast, $\lambda_1$ has the eigenvector
$\vec{\rho}_1 = (1,-1,0)$ and defines stability in the amplitudal
directions; $\lambda_1 >0$ indicates a break in the amplitudal
symmetry.  $\lambda_2$ is always negative and does not alter
stability.

\textbf{[$\bf{A_0}$]} Fixed points at which two polaritors have
distinct finite amplitudes and the same direction ($R_1\!>\!R_2\!>\!0$
and $\psi\!=\!0$; Fig. \ref{fig:Fig2}(a)). The condition of the fixed points is defined by
$R_1^2\!+\!R_2^2\!\!=\!\gamma \!+\!\delta$ and $R_1^4\!+\!R_2^4 \!+\!
\left(\gamma\! +\!\delta \right)^2 \!=\! 2 \mu R_1R_2$ (see
Eqs. \!\!(\ref{eq:R1/2}-\ref{eq:Psi2})). Two types of solutions appear
as a pair of saddle-node bifurcations at
$4(\gamma+\delta)^2\!+\!\mu^2\!=\!4$. One of them is always unstable,
irrespective of parameter values, and is never realized. Only the
other fixed point, denoted as $\rm{\bf{A_0}}$, can be realized. By
denoting $\gamma'\!  \equiv\!\gamma\!+\!\delta$, square amplitudes of
the solution are defined as
$R_1^2 \!=\! [ \gamma' +\! \{\gamma'^2-\!(-\mu-\! \sqrt{\mu^2\!+\!4\gamma'^2\!-\!4} )^2\}^{1/2}]/2$ and
$R_2^2 \!=\! [ \gamma' -\! \{\gamma'^2-\!(-\mu-\! \sqrt{\mu^2\!+\!4\gamma'^2\!-\!4})^2 \}^{1/2}]/2$.
Through the change in the amplitudal direction, this fixed point
connects to $\rm{\bf S_0}$ via pitch-fork bifurcation and connects to
$\rm{\bf P_n}$ via transcritical bifurcation
(Fig.~\ref{fig:Fig2}(a)). Angular stability is lost when the
eigenvalue $\lambda_3 \!=\! 2 \gamma' \delta \!-\! 2 \nu R_1R_2$ is
positive, with an eigenvector of $\vec{\rho}_3 \!=\!  (0,0,1)$.

\textbf{[$\bf{A_{\pi}}$]} Fixed points at which two polaritors have
distinct, finite amplitudes and opposite directions
($R_1\!>\!R_2\!>\!0$ and $\psi\!=\!\pi$; Fig.~\ref{fig:Fig2}(a)).  Similar to the case of
$\rm{\bf A_0}$, two types of fixed points appear as a pair of
saddle-node bifurcations at $4\gamma'^2\!+\!\mu^2\!=\!4$.  One of
these points is always unstable, and only the other, $\rm{\bf
  A_{\pi}}$, can be realized. The solution relates $\rm{\bf S_{\pi}}$
and $\rm{\bf P_n}$ via pitch-fork and transcritical bifurcations,
respectively (Fig.~\ref{fig:Fig2}(a)).  Angular stability is lost when
$\lambda_3 \!=\!  2\delta(\gamma+\!\delta) \!+\!  2 \nu R_1R_2$ is
positive. Square amplitudes at $\rm{\bf A_{\pi}}$ are defined as
$R_1^2 \!=\! [ \gamma' +\! \{\gamma'^2-\!(\mu-\!
\sqrt{\mu^2\!+\!4\gamma'^2\!-\!4} )^2\}^{1/2}]/2$ and $R_2^2 \!=\! [
\gamma' -\! \{\gamma'^2-\!(\mu-\! \sqrt{\mu^2\!+\!4\gamma'^2\!-\!4})^2
\}^{1/2}]/2$.

\textbf{[$\bf{A_a,A_{b}}$]} Fixed points with distinct, finite
amplitudes ($R_1\!>\!R_2\!>\!0$) and a $\psi$ that is neither $0$ nor
$\pi$ (Fig.~\ref{fig:Fig2}(a). These solutions are obtained from the following conditions
derived from Eqs. (\ref{eq:R1/2}-\ref{eq:Psi2}): $R_1^2+R_2^2 = \gamma
+ \delta \cos 2 \psi$, $R_1^4+R_2^4 + \left(\gamma +\delta \cos
2\psi\right)^2 = 2 \mu \cos \psi R_1R_2$, and $\delta
\left(R_1^2+R_2^2 \right)\cos \psi = \nu R_1R_2$.  The expression
giving the solutions is lengthy and of secondary importance for the
purpose of this work. We showed that there can be twelve types of
possible solutions. Two of them bifurcate from $\rm{\bf A_0}$ and
$\rm{\bf A_{\pi}}$ by the angular instabilities through pitch-fork
bifurcations. Let us denote them as $\rm{\bf A_{a}}$ and $\rm{\bf
  A_{b}}$, respectively. These two fixed points are separated, since
$R_1\!=\!  R_2\!=\!0$ at $\psi \!=\! \pm \pi/2$; $|\psi|$ is less than
$\pi/2$ for $\rm{\bf A_{a}}$ but larger than $\pi/2$ for $\rm{\bf
  A_{b}}$.  The other fixed points are mostly unstable; two of them
may appear in the limited parameter regions via saddle-node or
subcritical bifurcation (these fixed points are not discussed further
in this report).

Figure~\ref{fig:Fig2}(b) shows the phase
diagram against $c_2 = (\mu - \nu)/2 $ and $\delta$, with fixed values
of $\gamma = 0.7$ and $c_2 = 1.0$.  At the fixed points $\rm{\bf
  P_n}$, $\rm{\bf S_0}$, $\rm{\bf S_{s}}$, $\rm{\bf A_0}$, and
$\rm{\bf A_{\pi}}$, cells move directionally straight ($\dot{\Psi} =
0$), but they have different shapes. Examples of cell motion at
$\rm{\bf P_n}$ and $\rm{\bf S_s}$ are shown in Fig.~\ref{fig:Fig2}(c)
and (d), which are similar to polarized and keratocyte-like cell
motions, respectively.  On the other hand, $\dot{\Psi}$ is nonzero and
the cells show migration with a circular orbit at the fixed points
$\rm{\bf A_{a}}$ and $\rm{\bf A_{b}}$ (Fig.~\ref{fig:Fig2}(e)).  A
cell at fixed point $\rm{\bf S_{\pi}}$ is bipolar and does not exhibit
migration.

In addition to these fixed points, numerical simulations revealed
oscillatory dynamics in narrow parameter regions \cite{Caption2}.  The
cell in the oscillatory state exhibits a quasi-periodic Lissajous
orbit, as shown in Fig.~\ref{fig:Fig2} (f). Another type of motion
found by numerical simulation is repetitive right angle turns
(Fig.~\ref{fig:Fig2}(g)).  Similar types of motions were reported for
another model \cite{HiraiwaEPL2010}.

Taken together, many inner states appear on account of the competition
among polaritors, even in the simplest system of $N=2$.  Note that
Eq.~(\ref{eq:Psi2}) indicates that $\psi =0$ and $\psi =\pi$ are
separatrices in the phase space of the $N =2 $ system. In addition, $r
\equiv R_2 - R_2$ obeys the equation of the form $dr/dt = Q \times r$
from Eqs.~(\ref{eq:R1/2}) and (\ref{eq:R2/2}), where $Q$ is a function
of $(R_1, R_2, \psi)$, indicating that $R_1=R_2$ is an additional
separatrix in the system. Thus, the order in amplitudes of polaritors
defined by the initial condition is maintained in a noiseless
system. These separatrices constrain the dynamics of the $N=2$ system;
for example, zig-zag motion is forbidden.

\begin{figure}[bt]
\centering
  \resizebox{0.42\textwidth}{!}{%
    \includegraphics{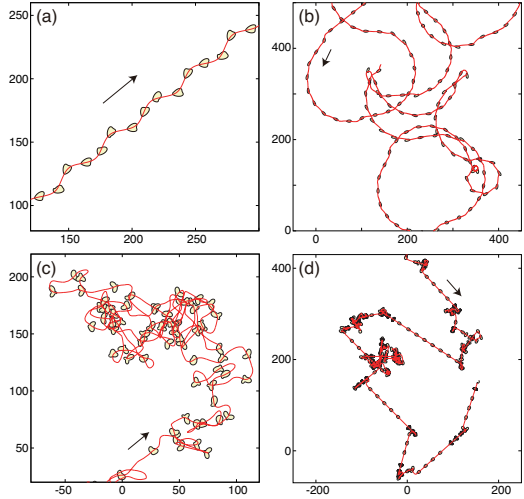}
  }
  \caption{\label{fig:Fig3} (a) Zig-zag motion for the $N=3$ system
    with parameters $(\gamma,\delta,c_1,c_2) =
    (0.78,0.3,0.6,0.8)$. (b) Chaotic motion for the $N=3$ system.
    Parameters are set as $(0.8,0.55,0.35,-0.8)$. (c,d) Examples of
    chaotic motions in the $N = 20$ systems. Parameters are set as
    $(0.7,0.3,0.6,0.8)$ (c) and $(1.2,-0.22,0.8,2.35)$ (d).
    Parameters for shapes are set as $(R_0, \Lambda, A) = (1.5, 12.0,
    0.3)$ for (a,b) and $(1.8,4.0,0.3)$ for (c,d).  }
\end{figure}

Such constraints are absent for $N>2$; therefore, the system can
exhibit various types of motion. For $N = 3$, zig-zag motion of cell
migration is observed, as shown in Fig.~\ref{fig:Fig3}(a), where two
oscillating polaritors dictate the position of the head of the cell
and periodically determine the direction of migration. Chaotic inner
dynamics is found in the case where the cellular trajectory is also a
chaotic orbit, as shown in Fig. \ref{fig:Fig3}(b).  For larger $N$
systems, additional types of dynamics appear.  Fig. \ref{fig:Fig3}(c)
shows an example of cellular motion at the $N = 20$ system, wherein
the cell shape can fluctuate significantly, similar to the case of
amoebic motion. Fig. \ref{fig:Fig3}(d) shows another example in the $N
= 20$ system, for which the cell exhibits repetitively straight
motion, followed by locally diffusive random migration.

\begin{figure}[tb]
\centering
  \resizebox{0.4\textwidth}{!}{%
    \includegraphics{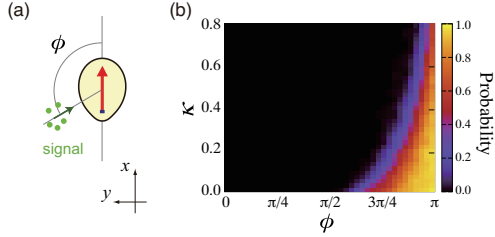}
  }
  \caption{\label{fig:Fig4} (a) Set-up of the numerical simulation. A
    cell with a polaritor directed in the $x$-direction receives a
    signal from direction $\phi$ at $t=0$.  (b) Probability of cell
    directional change by the replacement of polaritors is shown in a
    color scale against the invoked signal direction $\psi$ and
    parameter $\kappa$.  The replacement can occur only when $\phi >
    \pi/2$.  Parameters are set as $\gamma = 1.4$, $\delta = 0.15$,
    $\mu = 0.3$, $\nu = -0.1$, $\sigma = 2.0$, and $s= 1.0$. A small
    amount of noise is added during the simulation ($D=5.0 \times
    10^{-3}$).  The probability is calculated from 100 independent
    simulations for each parameter set.}
\end{figure}

Finally, we considered the chemotactic behavior of a cell in a signal
field $S(\mbox{\textbf{x}},t)$ \cite{Devreotes2003,VanHaastert2004}.
Cell shapes and internal cellular compasses are correlated with the
sensing ability to external signal molecules. Assuming that a cell can
sense the gradient of the signal field $s(\textbf{x}) \equiv \nabla
S(\textbf{x}) = s(\textbf{x})e^{i\phi(\textbf{x})}$, the coupling of
polaritor variables with the signal field is incorporated by combining
$d_1s |W_i|^2 + d_2 \overline{s} W^2_i$ with
Eq.~(\ref{eq:Wequation}). For $\sigma = d_1 +d_2$ and $\kappa =
d_1-d_2$, the following additional terms appear in Eqs. (\ref{eq:Rj})
and (\ref{eq:thetaj}):
\begin{eqnarray}
  \dot{R}^{sig}_i &=&  \sigma s R^2_i\cos(\theta_i-\phi), \label{eq:sigR}\\
  \dot{\theta}^{sig}_i &=& \kappa s R_i \sin (\phi-\theta_i).
\end{eqnarray}
Here, both $\sigma$ and $\kappa$ are set as positive for an attractant
signal. For a single polaritor with $R \sim 1$, the direction develops
as $\dot{\theta}_i \sim \kappa s \sin (\phi-\theta_i)$ and is directed
to the maximum gradient of the signal concentration, as introduced
previously \cite{HuPRE2010}. When $|\theta_i-\phi|$ is larger than
$\pi/2$, $\dot{R}_i^{sig}$ becomes negative and the existing polaritor
begins to shrink. Thus, the replacement of polaritors may occur,
depending on the direction of the signal gradient.

To confirm the response of cellular behaviors, numerical simulation is
conducted with the following settings. First, a cell with $N=2$
polaritors at the single polarity state $\bf{P_n}$ is prepared in the
absence of a signal field. The active polaritor is set in the
$x$-direction. Then, a signal field is applied at $t=0$ with
$\textbf{s}(t) = s e^{i\phi} \Theta(t)$, where $s$ and $\phi$ are
constants and $\Theta(\cdot)$ is a Heaviside function
(Fig.~\ref{fig:Fig4}(a)). Weak noise is added ($D=5.0 \times
10^{-3}$). Then, the system shows either rotation of the existing
polaritor (Fig.~\ref{fig:Fig1}(c), top) or the replacement of
polaritors (bottom), depending on signal direction $\phi$ and other
parameters.  Fig.~\ref{fig:Fig4}(b) shows the probability of the
replacement occurring for respective signal direction $\psi$ and
parameter $\kappa$.  The results demonstrate that the model cell can
respond to signal stimuli from the rear ($\phi > \pi/2$) by switching
polaritors.


By considering competition among polaritors, the present study
investigates a mechanism for organizing a variety of cellular
behaviors linked to morphology and migration.  The proposed model
exhibits distinct polar, keratocyte-like, zig-zag, and chaotic
amoeboid motions that are relevant to experimental observations. This
model is only constrained by the symmetry of the system. As
demonstrated in different systems like quadrupedal locomotion
\cite{Golubitsky}, such an approach based on symmetry can be helpful
in classifying a variety of possible cellular motions.  In addition,
the model is quite simple and requires little computational power,
making it possible to use the model to study collective cellular
behaviors \cite{Vicsek,LevinePRE2001}.

An advantage of our model is that it is easily extensible to higher dimensions
even with the same number of model parameters, and it provides an intuitive and
consistent interpretation of cellular behavior. Previous models have been reported
\cite{Ohta_PRL2009,HiraiwaEPL2010} that exhibit
similar dynamics to those presented in this study. However, these models become complicated by including higher-order tensor variables; in fact, zig-zag and chaotic motions appear in the equations that contain more than 20 parameters \cite{HiraiwaEPL2010}.

Because the concept of a polaritor is introduced here as a rather abstract variable, a future step will be to identify the molecular basis of the polaritors and their interactions. Validating Eq.~(\ref{eq:Wequation}) from
detailed subcellular processes (\textit{e.g.}, reduction from detailed
models
\cite{Otsuji2010,WeinerPLoSBiol2007,Arai20102012,Taniguchi,Jinken,Otsuji2007,Skpsky,Shao,Neilson2011,Nishimura,ShlomovitzPRL2007,Carlsson2010,DoubrovinskiPRL2011,EnculescuNJP2011,ZIebertJRSI2012})
will elucidate the way in which cellular motion depends on molecular parameters, which improves the correspondence of the model with experimental observations.

The author thanks D. Taniguchi, A. Nakajima, S. Sawai, and K. Kaneko
for the valuable comments. This work was supported by the Grant-in-Aid
MEXT/JSPS (No. 24115503).


\begin{thebibliography}{99}
\bibitem{Caption1}
    Interaction terms among three polaritors like $W_iW_j
    \overline{W}_k$ are ignored.
\bibitem{Caption2}
    No chaotic motion is found in the $N = 2$ system.
\end{thebibliography}
\end{document}